\documentclass[conference]{IEEEtran}
\IEEEoverridecommandlockouts
\usepackage{cite}
\usepackage{amsmath,amssymb,amsfonts}
\usepackage{algorithmic}
\usepackage{graphicx}
\usepackage{textcomp}
\usepackage{xcolor}
\def\BibTeX{{\rm B\kern-.05em{\sc i\kern-.025em b}\kern-.08em
    T\kern-.1667em\lower.7ex\hbox{E}\kern-.125emX}}
    
\usepackage[english]{babel}
\usepackage[utf8x]{inputenc}
\usepackage[T1]{fontenc}
\usepackage{amsthm}

\usepackage[bookmarks=false]{hyperref}
\usepackage{balance}  % for  \balance command ON LAST PAGE  (only there!)
\usepackage{booktabs} % For formal tables
\usepackage{amsmath,amssymb,amsfonts}
\usepackage{algorithmic}
\usepackage{graphicx}
\usepackage{textcomp}
\usepackage{color}
\usepackage{xcolor}
\usepackage[english]{babel}
\usepackage[utf8x]{inputenc}
\usepackage[T1]{fontenc}
\usepackage{amsmath}
\usepackage{graphicx}
\usepackage{listings}
\usepackage{url}
\usepackage{cite}
\colorlet{punct}{red!60!black}
\definecolor{background}{HTML}{EEEEEE}
\definecolor{delim}{RGB}{20,105,176}
\colorlet{numb}{magenta!60!black}

\definecolor{light-gray}{gray}{0.80}
\newcommand{\eat}[1]{}
\newcommand{\scream}[1]{{\bf * #1 *}{\typeout{#1}}}
\newcommand{\red}[1]{\textcolor{red}{#1}}
\newcommand{\cadd}{{\em AddCite}}
\newcommand{\cdel}{{\em DelCite}}
\newcommand{\cmod}{{\em ModifyCite}}

\newcommand{\cmerge}{{\em MergeCite}}
\newcommand{\ccopy}{{\em CopyCite}}
\newcommand{\cf}[1]{$\mathcal{C}$ $_{#1}$}
\newcommand{\ct}[2]{{$Cite_{#1}(#2)$}}

\lstdefinelanguage{json}{
    basicstyle=\ttfamily\small,%\normalfont\ttfamily,
    numbers=left,
    numberstyle=\scriptsize,
    stepnumber=1,
    numbersep=8pt,
    showstringspaces=false,
    breaklines=true,
    frame=lines,
    backgroundcolor=\color{background},
    literate=
      {:}{{{\color{delim}{:}}}}{1}
      {,}{{{\color{delim}{,}}}}{1}
      {\{}{{{\color{delim}{\{}}}}{1}
      {\}}{{{\color{delim}{\}}}}}{1}
      {[}{{{\color{delim}{[}}}}{1}
      {]}{{{\color{delim}{]}}}}{1},
}

\begin{document}

\title{Automating Software Citation using GitCite}

\author{\IEEEauthorblockN{Leshang Chen and Susan B. Davidson}
\IEEEauthorblockA{\textit{Dept. Computer and Information Science} \\
\textit{University of Pennsylvania}\\
% Philadelphia, USA \\
\{leshangc, susan\}@seas.upenn.edu}
% \and
% \and
% \and
%\and
%\and
}

\maketitle

\begin{abstract}
The ability to cite software and give credit to its authors and contributors is increasingly important. While the number of online open-source software repositories has grown rapidly over the past few years, few are being properly cited when used due to the difficulty of creating appropriate citations and the lack of automated techniques. This paper presents GitCite, a model for software citation with version control which enables citations to be inferred for any project component based on a small number of explicit citations attached to subdirectories/files, and an implementation that integrates with Git and GitHub. The implementation includes a browser extension and a local executable tool, which enable citations to be added/modified/deleted to software project repositories and managed through functions such as fork/merge/copy.  
% This document is a model and instructions for \LaTeX.
% This and the IEEEtran.cls file define the components of your paper [title, text, heads, etc.]. *CRITICAL: Do Not Use Symbols, Special Characters, Footnotes, 
% or Math in Paper Title or Abstract.
\end{abstract}

% \begin{IEEEkeywords}
% component, formatting, style, styling, insert
% \end{IEEEkeywords}

\section{Introduction}
\label{sec: intro}
%Importance of software citation
As software becomes an increasingly important research product in Big Data-driven science, appropriate citation is essential to motivate members of the community to continue to contribute as well as to enable reproducibility (see e.g., \cite{rml2017}).
Influential statements on the importance of software citation have been published within the digital libraries community (e.g. FORCE 11 reports \cite{FORCE11_2014, FORCE11-Software}). 
However, despite widespread agreement about the importance of these standards, their complexity and the effort entailed discourages users from following them when generating citations by hand.  
Systems must therefore be developed to automate citations and deliver them along with the software component being used.  

To address this problem, several {approaches} have emerged. 
First, citations may be explicitly embedded  in the comment section of the code in a repository by project members. This approach requires significant effort on the part of both project members and users of the code.  Second, a released version of a software project may be treated as open-access data and uploaded to a public hosting platform {(e.g., Zenodo~\cite{zenodo})} which provides a DOI, thus enabling a traditional form of citation and ensuring persistence. This approach still requires a citation to be constructed for the release version.
Third, plug-ins can be developed to take a code source URL as input and return a citation for the whole project as a string. This approach reduces effort on the part of both project members and users since metadata is automatically extracted from the repository to construct the citation, which is then automatically returned to the user {(e.g., the Software Citation Tool~\cite{mozillasoftwarecitation})}. 

However, there are several important aspects of software citation that none of the current {approaches} address, which we discuss next. 

\textbf{Granularity}: %The citation to a component of a project (e.g. a file) may be different from that for the project as a whole; there is a need for {\em fine-grained} citations. This is not supported by the last two solutions as the citation applies to the project as a whole.
{A software project has many different components.  It may include different versions, each of which is a large directory/tree structure. 
Different project versions may be managed and contributed to by different people, and different subtrees of a project version may involve different contributors. Therefore, the citation to a component of a project (e.g. a file) may be different from that for the whole project since the contributors are different; there is a need for {\em fine-grained} citations.}

{\textbf{Modularity:} While the first {approach} enables fine-grained citations, explicitly attaching a citation to the comment section of every file is cumbersome. There may also be a lot of repeated information, since source code files in the same folder typically share common metadata.
Therefore, the approach should scale to large repositories by reducing user effort and sharing common metadata across citations when possible.}

{\textbf{Usability:} Since project repositories are constantly evolving, a software citation system should enable citations to be updated as changes are made to files. The {current approaches} are static, and {most} treat source code as a published snapshot or webpage. Integrating a software citation tool within a (heavily used) version control system such as Git {will allow} citations to be versioned along with the code, and carried through known operators such as fork.  Citations may also be modified as contributions to the same file in different versions change over time.}

\begin{figure*}[htbp]
    \centering
    \includegraphics[width=0.8\textwidth]{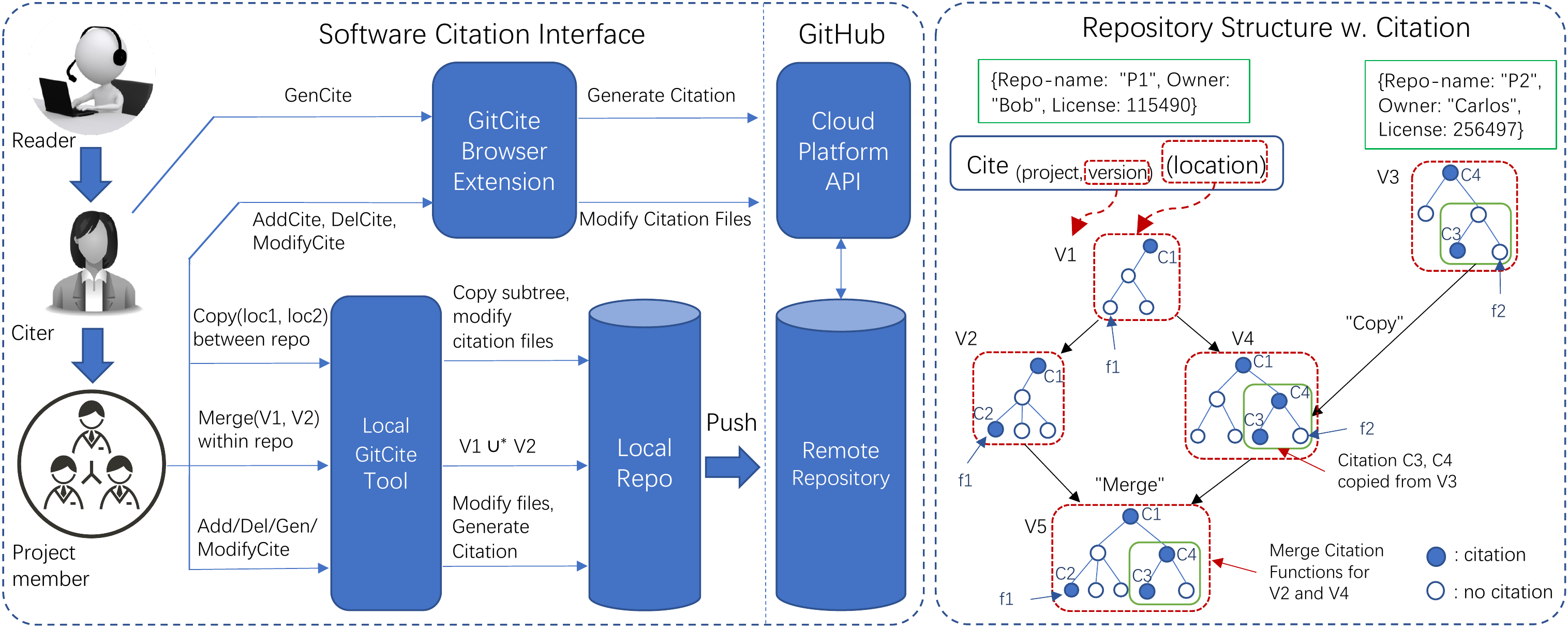}
    \vspace{-2mm}
    \caption{Architecture of GitCite (left) and Running Example (right).}
    \label{fig:systemarch}
    \vspace{-5mm}
\end{figure*}

%\red{\scream{Rewrote to clarify, not sure if all your points are still being made.}}

{
\textbf{Example.} 
To illustrate the need for fine-grained citations, we use {as an example} a project called {B} on GitHub  that is owned by {Bob}.  The {B} project imports a component called {CC} from a code repository called {C} {owned by Carlos}, and modifies {CC} to dovetail with other parts of the project. Although a comment appears in the readme file for {CC} giving credit to {Carlos}, it is not part of the metadata for the file and can easily be overlooked. 
%Furthermore, it is not clear what parts of the file are credited to Chen Li versus Yinjun Wu.  
If another user uses {CC} and cites the project repository as a whole, credit would wrongly be given to {Bob}. 
%Thus, it is necessary for different parts of the repository to have different citations, while the same part share similar ones. 
The {B} project code is then branched to enable another {contributor}, {Alice}, to independently develop a new component in a separate directory, which is later merged with the main branch of code development. 
Although the blame history reveals that {Alice} creates and edits files between versions, this can again be easily overlooked and credit incorrectly given to the repository owner {Bob}. \eat{\red{Thus, the three aspects above are necessary for code citation. }}
\eat{For other readers, however, it is hard to tell the independent contribution of Yanssie (especially when they ``Star'' the repository). When citations are generated by the URL, the credit will still be given to Yinjun, the repository owner. Thus,  citations should be able to be updated to reflect the version changes during the development stage of the project. }
}
\qed

%Technical Challenges
{\textbf{Challenges.} 
There are tensions between the granularity, modularity and usability requirements. Fine-grained citations are necessary to give accurate credit to authors, as the example shows. However, if the granularity is very fine (e.g. every line has its own citation), then usability and modularity may be impaired -- the workload on the user will increase and little metadata will be shared. 
In addition to usability issues, a line-level citation model is not compatible with a Git version-control model, and may introduce significant storage overhead. Our solution therefore balances these concerns, and considers citations at the level of files and directories. It also separates citations from code, enabling modularity and usability.
}

%Overview of our system, outline of demo paper.
{\textbf{Contributions.} 
This paper presents a model for software citation with version control based on a notion of a {\em citation function} (Section~\ref{sec: model}).  The citation function is designed at the level of files and directories, although it could be extended to the level of subsets of lines.
Citations can be specified for a subset of nodes in a directory structure and inherited by other nodes in the directory; the citation is also under version control. We show an interface called GitCite based on our model which is built on top of Git~\cite{torvalds2010git} and GitHub~\cite{github} (Section~\ref{sec: impl}). GitCite allows users to manage citations in their own repositories and ``inherit'' citations from other repositories. \eat{ 
The architecture of GitCite is shown in Figure \ref{fig
:systemarch} (left) \red{and discussed in Section~\ref{sec: impl}}. }}
In the demo (Section~\ref{sec: demo}), we will first give an overview of our citation system and introduce the main components.
Next, we will present a short video to demonstrate how the  citation system is used. Finally, attendees will be able to interact with the system to obtain, create, modify, and delete citations.

{\textbf{Other Related Works.} In software engineering community, some \textit{package management} tools like PyPI and Maven have been made and the dependency metadata of the tools is used to track the popularity of repositories \cite{githubdependency}. Some other tools like git-subtree and git-submodule \cite{gitsubtree} focus on managing partially organized source code. These approaches typically aim at setting up software environment while citations \cite{DBLP:conf/jcdl/AlawiniCDSS17} focus on claiming authorship and giving credit.  %In software engineering community, some \textit{package management} tools like PyPI \cite{pip} and Maven\cite{maven} have been made. GitHub has started to utilize the metadata from the tools to track the popularity of repositories based on dependency \cite{githubdependency}. Some other approaches like git-subtree and git-submodule \cite{gitsubtree} focus on managing partially organized source code. These approaches typically aim on setting up software environment while citations focus on claiming authorship and giving credit. Also, we do not address {\em persistance} in this paper, which is partially solved by \eat{projects like Zenodo~\cite{zenodo} and }the Software Heritage project~\cite{heritage} and can be integrated with our solution. 
}

\eat{
PUT THIS IN THE CONCLUSIONS ALONG WITH FUTURE WORK?
It should also be noted that including such citation information does not necessarily guarantee {\em persistance}, which is an important component of the FORCE 11 reports~\cite{FORCE11_2104,FORCE11-Software}.  
Code repositories on public platforms like GitHub can therefore be archived to services such as Zenodo~\cite{zenodo},
which provides a DOI that can be embedded in a reference along with other citation information, thereby addressing the issue of persistance. 
However, apart from copy-pasting citations from the headers of files (when such information exists), there is no automatic way to generate citations to software retrieved from software repositories.  

Should also discuss bibliometrics over the software.
Should we discuss history and Wikipedia issues?
}

\vspace{-1mm}
%\subsection{Use Cases}
\section{Citation Model}
\label{sec: model}
{We %introduce the notion of a project repository before describing  
now define the notion of a citation function, describe how it is modified as a result of various operations  (add/delete/modify citations, copy/fork/merge subtrees), and discuss how this addresses our technical challenges.} %Removed the following data citation references: % Note that this model is strongly influenced by ideas in~\cite{BunemanEtAl2016, BunemanSilvello2010}.

\textbf{Roles of Users.} To clarify the capabilities of users within the citation system, we start by describing the roles of users. Citation involves three types of users: {project member}, {citer}, and {reader}.  A {\em project member} can edit a project repository, issue Git operations,
%commands (commit, merge, pull requests, etc.), 
and manage citations attached to project components (directories or files). A {\em citer} requests the citation for some component of a project (called GenCite), and may inject the citation into a versioned or unversioned research product (article, paper, website or other code repository). A {\em reader} sees a citation and may dereference it to obtain the software being cited. %A user may have multiple roles, depending on her character and operation in the project. 

\begin{figure*}[t]
\centering
\includegraphics[width=0.75\textwidth]{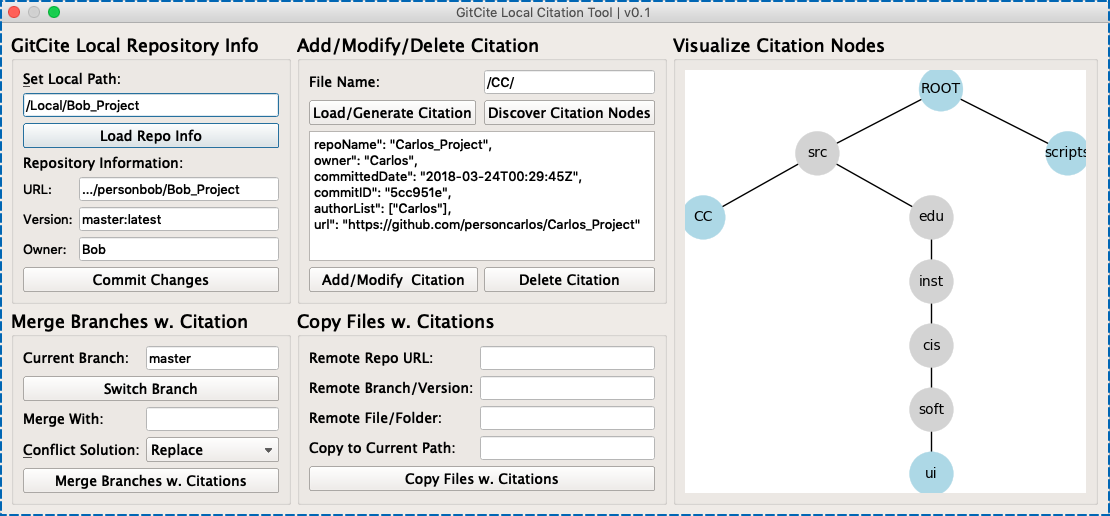}
\vspace{-2mm}
\caption{GitCite \eat{. (a) Browser Extension. (b)}Local Executable Tool. }
\label{fig:localexecutable}
\vspace{-5mm}
\end{figure*}

%\noindent
{\bf Citation function.} 
A project repository is a directed acyclic graph of project {\em versions}.  Each version is a directory, i.e. a rooted tree whose interior nodes are directories and leaves are files.  
{To address granularity, we allow citations to be defined for any node in any version of a repository.} 
%\scream{I thought we could  only defined citations on the \textbf{current} version, not past versions?}
Each version $V$ in project $P$ has an associated {\em citation function} \mbox{\cf{(V, P)}{}} which, given a path in the tree, returns a citation for the node at the end of the path. \eat{The function may be partial, i.e. some of the directories/files may not have explicit citations attached. } The root of each version \textit{must} be in the active domain, and its citation includes basic snippets of information such as the owner and name of the repository, the http address or DOI of the version, and the version number and/or date.

{To address modularity, the citation function may be partial, i.e. some of the directories/files may not have explicit citations attached.  Nodes with no explicit citations may inherit the citation from other nodes (defined in a policy).} In this demo, we define the citation of a node as follows:  Given a node $n$ in version $V$ of project $P$ with root $r$, the citation of $n$, denoted \ct{(V,P)}{n}, is defined as follows:
% \eat{
% \begin{itemize}
%     \item \cf{(V,P)}($n$) $\cup$ \cf{(V,P)}($r$), if \cf{(V,P)}($n$) is defined
%     \item \cf{(V,P)}($a$) $\cup$ \cf{(V,P)}($r$), where $a$ is the closest ancestor of $n$ for which a citation is defined
% \end{itemize}
% }
\begin{itemize}
    \item \cf{(V,P)}($n$), if $n$ has a citation defined (i.e. $n$ is in the active domain of the citation function)
    \item \cf{(V,P)}($a$), where $a$ is the closest ancestor of $n$ for which a citation is defined
\end{itemize}
Since the root must be in the active domain, \cf{(V,P)}($n$) is defined for every node. 

%there could be other definitions of \ct{(V,P)}{n}, e.g. ones that include every citation on the path from $n$ to $r$.

{\bf Adding/deleting/modifying citations.} {For usability, it should be possible to update citations to reflect the development of the source code repository.} As a project version is modified by inserting/deleting/modifying or renaming files, project members may also modify its citation function by adding (\cadd), deleting (\cdel), or modifying (\cmod) citations.  Each of these operators takes as input the path of the file/directory whose citation is being modified;  \cadd\ and \cmod\ additionally take the value for the new or modified citation.  
%The interpretation of these operators is straightforward.  
These operations may also be applied by the system as a side-effect of {Git} operations on the repository. %, which we discuss next. 
%\scream{This is not clear, see how I reworded.  Can't there be multiple "current versions"?  and can't you fork from a past version?  Maybe remove this?}
{Note that in version control systems such as Git, updates to the project are append only.  Therefore, GitCite allows citation updates only on the latest version of a branch.}

\eat{When a new version $V'$ is created, \cf{(V',P)} is created by applying these updates to \cf{(V,P)}.  The citation to the root of $V'$ will automatically be updated with the new version number. We assume that snippets such as http, DOI, version number and date cannot be directly modified by users.}

\eat{
\scream{Possibly delete this}
Modifications to files/directories and to their associated citations are \textit{independent}, i.e. the citation for a file could change but the file remain unchanged between versions, or the file could change but the citation remain unchanged between versions.  However, if a file or directory in the active domain of the citation function is moved or renamed  %from one directory to another within the project, 
then the citation function must be modified to reflect the file or directory's path in the new version.
}

% \vspace{-0.5mm}
{\bf Carrying citations through Git operators.}
{Another aspect of usability is to keep citations attached to files and directories as they are moved.} For example, branches may occur in a repository by separate edits being made to the same version. At some point, two branches $V1$ and $V2$ may be {\em merged}.
%, in which case the citation functions \cf{V1} and \cf{V2} must also be merged (\cmerge). 
%possibly move some of the below to next section as well.
\eat{
Git can resolve simple cases. For example, if one file is deleted in branch $V1$ and present in branch $V2$, then Git would do automatic merge based on the timestamps of commits. In the simplest cases, Git can also be used to resolve the conflicts of citations based on the way of saving it.  However, there is no guarantee that the code-based conflict resolution of Git will always fit into the key-value pair of citations. Thus, in order to be safe, such kind of conflict needs to be handled in addition. When conflict happens, users should be able to choose either version of citation to keep consistency and reflect human idea. When citation functions are simply added to both branches without conflict,  all of the added citations should be combined together. }
%\scream{[Leshang Note: which is the best place to talk about the conflict resolution. Here or in Sec 3? ] Leshang:  Discuss this further in the text to align with what you have implemented.  If one branch deletes a file and the other doesn't, what happens?  We want to make sure that the domain of the citation functions remains the active nodes in the directory.}
A subdirectory $T$ may also be {\em copied} from version $V1$ of repository $P1$ to version $V2$ of repository $P2$,
or a new project $P3$ may be {\em forked} from the current version of a project $P1$. % (\ccopy).  
Through each of these operations, the citation function associated with the new version must be made {\em consistent} with new directory structure and the files retained in the new version. % , i.e. the domain of the function must be consistent with the new version.  
It should also be as {\em complete} as possible (modulo conflicts).  
\eat{{\color{red} All of the above citation functions are mostly used by project member.} }
We will discuss how we implement {\em MergeCite}, {\em CopyCite} and {\em ForkCite} in more detail in the next section. 

\eat{In this case, the citations for each node in $T$ are augmented with the root citation for $V1$ and added to \cf{(V2,P2)}.  That is,  if $r$ is the root of $V1$, then for each node $n$ in $T$, 
\cf{(V1,P1)}($n$) $\cup$ \cf{(V1,P1)}($r$) is added to \cf{(V2,P2)}. If \cf{(V1,P1)}($n$) is not defined, then \cf{(V1,P1)}($a$) is used instead. 
}

% \vspace{-0.5mm}
{\bf Example:}  The right half of Figure \ref{fig:systemarch} illustrates these ideas. Each red-dotted box shows a version of a repository. In particular, $V3$ is a version of project $P2$ while the others are versions of project $P1$. The tree structure in each box shows the directory structure for the repository at a commit point. A solid blue circle means the node (file or directory) has an attached citation and the blue rimmed circle means that there is no attached citation. All versions have a default citation attached to the root. 
From $V1$ to $V2$, an {\cadd} operation attaches a citation to the leftmost leaf node, $f1$. 
%Based on closest ancestor definition, the citation for a folder can also serve for a file under that folder, if it doesn't have citation attached. 
Therefore, before adding the citation \ct{(V1,P1)}{$f1$}=$C1$, whereas afterwards \ct{(V2, P1)}{$f1$}=$C2$. 
\eat{
When the subtree of $V3$ in the green box is copied from project $P2$ to $P1$ by {\ccopy}, the attached citation is also copied along with the files. {\ccopy} would also change the citation. Before copying, \ct{(V3, P2)}{$f2$}=$C4$. After copying to $P1$, \ct{(V4, P1)}{$f2$}=$C4 \cup C1$.  At last, $V2$ is combined with $V4$ by {\cmerge} to produce $V5$, in which the citations are also merged along with the files. In this example there is no complex conflict when merging, so citations are simply combined. \ct{(V5,P1)}{$f1$}=$C2 \cup C1$. 
}

\vspace{-1mm}
\section{Architecture}
\label{sec: impl}

% \begin{figure}[htbp]
% \centering
%     \begin{subfigure}[t]{0.4\textwidth}
%     \centering
%     \includegraphics[width=\textwidth]{chromeextension2.png}
%     \end{subfigure}
%     ~ \\
%     \begin{subfigure}[t]{0.4\textwidth}
%         \centering
%         \includegraphics[width=\textwidth]{cmdlinetool.png}
%         %\caption{Buyer GUI}
%     \end{subfigure}
%     \caption{GitCite tools. Top: browser extension; Bottom: local executable command line tool}
%     \label{fig:chromeextension}
% \end{figure}

We now describe the architecture of our citation system (see left half of Figure \ref{fig:systemarch}). 
%We provide a basic solution for citing code in Git repositories managed by source code versioning system and hosted on remote platform like GitHub. Figure \ref{fig:systemarch} shows the overall architecture. 
CitCite consists of two components: a  {\em browser extension}, and a {\em local executable tool}.
%the citation extension tools for Git. 
Together they provide a citation service for GitHub software repositories.
Since the local executable tool is based on Git, it is also compatible with any other online project management website which uses Git.  
%that are maintained by Git. 
%These two tools can be installed on local computer easily and can be used with great convenience. 
We discuss how citation functions are stored before describing the components.

% \begin{figure}[t!]
% \centering
% \includegraphics[width=0.45\textwidth]{gitcitationlocalgui2.png}
% \caption{GitCite Local Executable Tool}
% \label{fig:localexecutable}
% \end{figure}

% \begin{figure}[t!]
% \centering
%     \begin{subfigure}[t]{0.45\textwidth}
%     \centering
%     \includegraphics[width=\textwidth]{citationextensionICDE3.png}
%     \end{subfigure}
%     ~ \\
%     % ~ \\
%     \begin{subfigure}[t]{0.45\textwidth}
%         \centering
%         \includegraphics[width=\textwidth]{gitcitationlocalgui.png}
%         %\caption{Buyer GUI}
%     \end{subfigure}
%     \caption{GitCite tools. Top: browser extension; Bottom: local executable command line tool}
%     \label{fig:chromeextension}
% \end{figure}

%\scream{We haven't discussed persistence, the red scream part doesn't really make sense.}
% \vspace{-0.5mm}
%\subsection
{\bf Storing Citation Functions.} To {address  modularity and usability, we store citations in a special {citation file} at the root of each version of a Git repository.} The file is a set of key-value entries, where the key is the relative path to the file being cited, and the value is the citation attached to the file. It cannot be directly modified by users, but is updated as a side-effect of the various citation and Git operations.  When a new commit version of the project is created, the updated citation file is also created which reflects the updates made since the previous version.  
These updates include adding/deleting/modifying entries as a result of \cadd, \cdel, or \cmod\ operations; changing the key of an entry to reflect a renamed file or directory; or adding new entries in response to copying/merging between branches. 
Thus, much of the work involved in storing and maintaining citations is done as a side-effect of versioning in Git. 

%Thus, in order to use the citation function, the developer should use our tools each time they want to change the citation for the repository. 

\eat{

{\color{red} \textbf{Scalability.} We consider the scalability of citation to be two-fold: across platforms and inside a repository. %When considering large amount of repositories and copies to be stored everywhere, we need the citation to be fetched and processed efficiently. 
Since our implementation stores citation file inside Git repository, we naturally transfer the scale problem to distributed version control system. Since Git is well designed for distributed collaboration, we would definitely get the content dispatched and synchronized among different locations at real time. Inside one repository, since we store citation in the root, and follows closest-ancestor coverage, we reduce the number of citations to be written in the file and there's no need to attach citation to each file when multiple citations under a tree are the same. The time of finding closest ancestor, which affects the latency, is negligible under careful design of string prefix comparison algorithm. 
}}

\begin{figure}[t]
\centering
\includegraphics[width=0.45\textwidth]{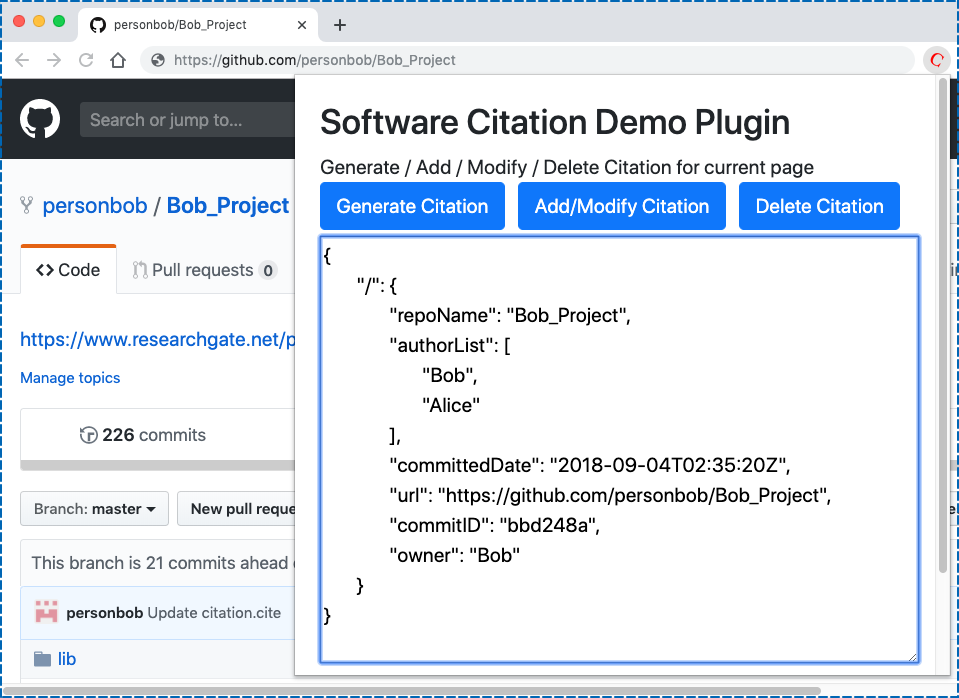}
\vspace{-2mm}
\caption{GitCite Browser Extension. }
\label{fig:chromeextension}
\vspace{-5mm}
\end{figure}

% \vspace{-0.5mm}
%\subsection
{\bf Browser extension.}
%\scream{Note: Modified. Generate is confusing since we have not discussed. Also it is not clear how the functionality changes between a "user" (someone who is not credentialed to make changes to citations) and a "member" (someone who is).  See if what I say is correct! }
The GitCite browser extension is deployed on Chrome\eat{ and written in JavaScript}, and can generate citations for remote repositories on GitHub.  The extension communicates with the GitHub server using its\eat{ REST} API, and directly modifies the citation file on the %currently open 
remote repository to reflect changes to the citation function.  
The popup page shown in Figure \ref{fig:chromeextension}
illustrates its functionality:  \eat{Users provide their credentials on GitHub to obtain access to the repository, and may then click on a node %(directory or file) 
and use the browser extension. }%{\em Generate} citation implements \ct{(V,P)}{n}, and starts by {\em populating} the citation.   under different conditions of user permissions and existence of citation file. 
If the user is a citer, clicking on ``Generate Citation'' will immediately generate the citation based on the {citation file} in that version; recall that this is either the citation explicitly attached to the node or that of its closest ancestor.  The returned citation can then be copy-pasted to their local bibliography manager. Since citers are not allowed to add/delete/modify files, they will not be allowed to use the ``Add/Delete'' button functionalities.  %Note that every node will have {\em some} citation by definition, even if it is the default citation constructed based on GitHub metadata.  
However, if the user is a project member, the text box will display the citation explicitly attached to the node, if it exists, which the user may then modify.  If an explicit citation does not exist, the user may either enter a citation, or use the ``Generate\eat{ Citation}'' button to see the citation of its closest ancestor, which can then be modified for the current node. If no citation has ever been enabled for the repository (as  is the case for most existing repositories), then a default citation would be {created} from the metadata of the GitHub repository\eat{, similar to \cite{mozillasoftwarecitation}}, {and the project member could use it as a draft that they could further edit before saving.} 
 
%\subsection
{\bf Local executable tool.}
When a project member downloads a copy of the project repository with Git, the GitCite local executable tool (Figure \ref{fig:localexecutable}) can be used to manage the citation file in the download. In addition to \cadd, \cdel, and \cmod, it also implements the {\em CopyCite}, {\em MergeCite} and {\em ForkCite} functions. {\em CopyCite} copies a directory from a remote repository version to the local repository version, and {\em migrates} their associated citations. That is, the citations for the directory and its subtree in the remote {citation file}\eat{``citation.cite" file} are added to the local {citation file}, with the key paths modified to reflect the new location to ensure correctness of the citation function. % in the local repository version.  
%If the folder is copied to another folder under new repository, a prefix will be added to the key of the citations. If user requires the default citation to apply to all copied files, the closest citations matched with the copied files will also be copied, even if they are not designated for them. 
%\scream{Leshang Note: Modified. This is a good question. Another key question as mentioned in the model section: Default citation (closest matches) of V2 should be copied to V1 if the copied files in V2 does not have specified citation. Currently in implementation I just copy the citation exactly matches those files. I need to update the code. Original Note: Does it update keys?  i.e. the root-to-node path in the other repository is different than in the new repository.}
{\em MergeCite} merges two branches in the same repository, and merges the citation files while resolving conflicts. 
\eat{The types of conflicts include but are not limited to: Deleting a file that has an attached citation in one branch while keeping the file and attached citation in the other branch;
%, in which case the file and citation are deleted in the merge;
and changing the citation for the same file one or both branches. }
%This tool modifies files on local machine and deals with the citations attached to those modified files. 
Although Git conflict resolution rules are used for all regular files,
we do not use them on the citation file since it could leave the citation function inconsistent.  Instead, we simply take the union of the citation files, and delete any entries that correspond to files that were deleted by the Git merge. Conflicts over the values associated with the same key in the new {citation file} are then resolved by showing them to the user and asking the user to resolve the conflict.  
%Citation conflict during merging are firstly resolved using default rules of our tool, since the existing conflict helper of Git only solve the merge problem at line level which would destroy the citation. We provide a naive solution here:
%Merge those citations whose keys do not conflict between the branches. If any two citations with the same key conflict with each other, the citation of the main branch that starts the merge is kept in the new version and another is omitted by default. User also has the choice of selecting either version manually based on their preference. The naive merge leaves the user with other potential conflicts of files to resolve, and won't commit before user make additional modifications that justify the merged files and citations. 
%, it is better than using Git to merge the citation files, as our tool will guarantee the valid format of citation files and reduce the risk of corrupting the citations. Smarter way of synchronizing user's preference on keeping which version of files and which version of citations will be studied in the future. 
More complex conflict resolution strategies could also be used.
{\em ForkCite} copies a version of a repository, along with its history, and creates a new repository.  The citations in {the citation file} are also copied. %, and are not modified unless the user explicitly chooses to do so.
Our way of storing citations will naturally enable {\em ForkCite} through GitHub's {\em Fork}. 
{Users can also view the nodes with explicit citations attached in light blue on the right side of the tool. } When changes to files and their citations are finally committed, the Git push command is used to push the local copy (which contains {the citation file}) to the remote repository. 

{\bf Example, cont.}
Returning to the right half of Figure \ref{fig:systemarch}, when the subtree of $V3$ in the green box is copied from project $P2$ to $P1$ by {\ccopy}, the attached citations are migrated as described earlier. Since $f2$ does not have an associated citation in $V3$, {before the copy event} \ct{(V3, P2)}{$f2$}=$C4$, which is the citation of its closest ancestor.
Note that $C4$ is also the citation for the root of the subtree being copied.  After copying to $P1$ the citation for $f2$ remains unchanged:  \ct{(V4, P1)}{$f2$}=$C4$ since the citation for the root of the green subtree from $V3$ has been added to the citation file for $V4$, indicated by the root now being a solid blue color.  Finally, $V2$ and $V4$ are merged using {\cmerge} to produce $V5$, in which the citations are also merged along with the files. In this example there are no conflicts, so we simply take the union of the citation files.

\vspace{-0.5mm}
\section{Demonstration Scenario}
\label{sec: demo}
% \scream{Modify the demo scene}
During the demo, we will talk about the need for fine-grained software citation, explain our citation model, and show the GitCite architecture. {We will then demonstrate both the browser extension and the local executable functionalities of GitCite, showing how citations are managed through changes as well as how they can be added or modified within the current repository. Using {the \eat{(real world) }example in the introduction}, we will illustrate how our tools can be used to establish citations for the whole repository, such that different citations will be returned for different components.}
After this introduction, attendees will be invited to test our citation tools using the demo machine, either assuming the role of a citer to obtain citations from repositories of their choosing, or assuming the role of a project member to  generate/modify citations, or to copy/merge citations among our sample project repositories.

\vspace{-0.5mm}
\section{Conclusion}
This demo presents a model and implementation called GitCite for citing software project repositories managed by version control systems like Git. The system consists of a browser extension that works with GitHub, and a local executable tool that works with any project management platform using Git.  Citations may be explicitly attached to any subset of the nodes {to show its identity and authorship}, and the root of the project must always have a default citation.  The citation for a node is based on its {\em closest ancestor} with an explicit citation.  Users are able to create, manage and use citations through  {\cadd}, {\cdel}, and {\cmod} functions, as well as the citation-extended Git functions {\ccopy}, {\cmerge} and {\em ForkCite}. 
\vspace{-1mm}

% Several interesting problems remain.  First, we would like to explore other citation conflict resolution strategies beyond the simple union interpretation used in this demo, in particular ones that mirror the three-way merge method used in Git.  Second, since many software repositories have already been developed without being ``citation-enabled", we would like to explore ways of adding retroactive citations and ensuring their consistency and preservation through the project history. Third, we would like to see how to integrate our system with software archives such as the Software Heritage archive~\cite{heritage, DBLP:journals/cacm/AbramaticCZ18}.
%https://www.softwareheritage.org/archive/

% \balance

\bibliographystyle{IEEEtran}
\bibliography{icde_camera_ready}

\begin{thebibliography}{10}
\providecommand{\url}[1]{#1}
\csname url@samestyle\endcsname
\providecommand{\newblock}{\relax}
\providecommand{\bibinfo}[2]{#2}
\providecommand{\BIBentrySTDinterwordspacing}{\spaceskip=0pt\relax}
\providecommand{\BIBentryALTinterwordstretchfactor}{4}
\providecommand{\BIBentryALTinterwordspacing}{\spaceskip=\fontdimen2\font plus
\BIBentryALTinterwordstretchfactor\fontdimen3\font minus
  \fontdimen4\font\relax}
\providecommand{\BIBforeignlanguage}[2]{{%
\expandafter\ifx\csname l@#1\endcsname\relax
\typeout{** WARNING: IEEEtran.bst: No hyphenation pattern has been}%
\typeout{** loaded for the language `#1'. Using the pattern for}%
\typeout{** the default language instead.}%
\else
\language=\csname l@#1\endcsname
\fi
#2}}
\providecommand{\BIBdecl}{\relax}
\BIBdecl

\bibitem{rml2017}
{Panel Summary of Reproducibility in ML Workshop, ICML'17}.
  \url{https://sites.google.com/view/icml-reproducibility-workshop/icml2017}.

\bibitem{FORCE11_2014}
M.~Martone, Ed., \emph{{Data Citation Synthesis Group: Joint Declaration of
  Data Citation Principles}}.\hskip 1em plus 0.5em minus 0.4em\relax FORCE11,
  2014.

\bibitem{FORCE11-Software}
A.~Smith, D.~Katz, K.~Niemeyer, and {FORCE11 Software Citation Group},
  ``Software citation principles,'' \emph{PeerJ Computer Science}, 2016.

\bibitem{zenodo}
{Zenodo}. \url{https://zenodo.org}.

\bibitem{mozillasoftwarecitation}
{The Mozilla Science Team}. (2016) {FORCE11 Software Citation Tools}.
  \url{https://github.com/mozillascience/software-citation-tools}.

\bibitem{torvalds2010git}
L.~Torvalds and J.~Hamano. (2010) {Git}. \url{http://git-scm.com}.

\bibitem{github}
{GitHub}. \url{https://github.com/}.

\bibitem{githubdependency}
{Listing the packages that a repository depends on}. \url{
  https://help.github.com/en/github/visualizing-repository-data-with-graphs/listing-the-packages-that-a-repository-depends-on}.

\bibitem{gitsubtree}
{Git Tools - Subtree Merging}. \url{
  https://git-scm.com/book/en/v1/Git-Tools-Subtree-Merging}.

\bibitem{DBLP:conf/jcdl/AlawiniCDSS17}
A.~Alawini, L.~Chen, S.~B. Davidson, N.~P.~D. Silva, and G.~Silvello,
  ``Automating data citation: The eagle-i experience,'' in \emph{JCDL}, 2017.

\end{thebibliography}
% \bibitem{b1} G. Eason, B. Noble, and I. N. Sneddon, ``On certain integrals of Lipschitz-Hankel type involving products of Bessel functions,'' Phil. Trans. Roy. Soc. London, vol. A247, pp. 529--551, April 1955.
% \bibitem{b2} J. Clerk Maxwell, A Treatise on Electricity and Magnetism, 3rd ed., vol. 2. Oxford: Clarendon, 1892, pp.68--73.
% \bibitem{b3} I. S. Jacobs and C. P. Bean, ``Fine particles, thin films and exchange anisotropy,'' in Magnetism, vol. III, G. T. Rado and H. Suhl, Eds. New York: Academic, 1963, pp. 271--350.
% \bibitem{b4} K. Elissa, ``Title of paper if known,'' unpublished.
% \bibitem{b5} R. Nicole, ``Title of paper with only first word capitalized,'' J. Name Stand. Abbrev., in press.
% \bibitem{b6} Y. Yorozu, M. Hirano, K. Oka, and Y. Tagawa, ``Electron spectroscopy studies on magneto-optical media and plastic substrate interface,'' IEEE Transl. J. Magn. Japan, vol. 2, pp. 740--741, August 1987 [Digests 9th Annual Conf. Magnetics Japan, p. 301, 1982].
% \bibitem{b7} M. Young, The Technical Writer's Handbook. Mill Valley, CA: University Science, 1989.
% \end{thebibliography}
% \vspace{12pt}
% \color{red}
% IEEE conference templates contain guidance text for composing and formatting conference papers. Please ensure that all template text is removed from your conference paper prior to submission to the conference. Failure to remove the template text from your paper may result in your paper not being published.

\end{document}